\documentclass[aps,amsmath,pra,showpacs,twocolumn,groupedaddress]{revtex4}

\usepackage{graphics}

\newcommand \beq{\begin{equation}}
\newcommand \beqa{\begin{eqnarray}}
\newcommand \beqann{\begin{eqnarray*}}
\newcommand \eeq{\end{equation}}
\newcommand \eeqa{\end{eqnarray}}
\newcommand \eeqann{\end{eqnarray*}}

\begin{document}

\title{Reversal of the
circulation of a vortex by quantum tunneling in trapped Bose systems}

\author{Gentaro Watanabe$^{a,b,c,d}$ and C. J. Pethick$^{a,b}$}
\affiliation{
$^{a}$The Niels Bohr Institute, Blegdamsvej 17, DK-2100 Copenhagen \O,
Denmark
\\
$^{b}$Nordita, Roslagstullsbacken 23,
106 91 Stockholm,
Sweden\\
$^{c}$CNR-INFM BEC Center, Department of Physics, University of Trento,
Via Sommarive 14, 38050 Povo (TN), Italy
\\
$^{d}$The Institute of Chemical and Physical Research (RIKEN), 2-1
Hirosawa, Wako, Saitama 351-0198, Japan}

\date{\today}

\begin{abstract}

We study the quantum dynamics of a model for a vortex in a Bose gas with
repulsive interactions in an anisotropic, harmonic trap.  By solving the
Schr\"odinger equation numerically, we show that the circulation of the
vortex can undergo periodic reversals by quantum-mechanical tunneling.
With increasing interaction strength or particle number, vortices become
increasingly stable, and the period for reversals increases.  Tunneling
between vortex and antivortex states is shown to be described to a good
approximation by a superposition of
vortex and antivortex states (a Schr\"odinger cat state), 
rather than the mean-field state, and we
derive an analytical expression for the oscillation period. The problem
is shown to be equivalent to that of the two-site Bose Hubbard
model with attractive interactions.

\end{abstract}


\maketitle

Over the past decade, atomic Bose-Einstein
condensates have provided unprecedented opportunities for studying in
detail the properties of quantized vortices, both theoretically
\cite{fetter} and experimentally \cite{dalibard}.  Most theoretical
work has been based on the use of the mean-field, 
Gross-Pitaevskii (GP) approximation \cite{GPpapers}, and an interesting
prediction by Garc\'ia-Ripoll {\it et al.}  \cite{garcia-ripoll}
within this approach is that a rotating Bose-Einstein condensate in a
non-rotating anisotropic harmonic trap can undergo periodic reversals
of the sign of the vorticity if the initial energy of the system is
sufficient to overcome the energy barrier between vortex states with
opposite circulation.

Recent experimental developments make it possible
to realize few body systems trapped on an optical lattice
\cite{campbell,foelling}.
The behavior of such small systems can be quite different from what is
predicted from a
classical treatment based on the GP approximation.
In this paper we consider the problem of stability of a vortex
quantum-mechanically, and show by solving the Schr\"odinger equation
numerically that for energies below the barrier, reversals of the
vorticity can occur by tunneling.  We find that the wave function
during reversals resembles more closely 
a quantum superposition of states (a Schr\"odinger cat state) than
a mean-field one, and we derive an analytical expression for the rate
of reversals that agrees well with the numerical data.
Mathematically, the problem is equivalent to that of particles with
attractive interactions in a double-well potential, which has been
studied previously \cite{flach, ho}.

Consider $N$ identical bosons of mass $m$ in an anisotropic, harmonic
two-dimensional trap, with trap frequencies denoted by $\omega_x$ and
$\omega_y$, and for definiteness we shall assume that $\omega_x >
\omega_y$.  Following Ref.\ \cite{garcia-ripoll} we shall assume that
the only oscillator levels occupied are the first excited states of the
oscillators \cite{note_anharmonicity}, corresponding to the wave functions
$\psi_x\equiv C\
(x/d_x) \exp{\left(-x^2/2d_x^2-y^2/2d_y^2\right)}$, where
$C=\sqrt{2/(\pi d_xd_y)}$ and $d_x\equiv\sqrt{\hbar/m\omega_x}$ and a
similar expression for $\psi_y$.
The many-body Hamiltonian is
\begin{eqnarray}
H&=&H_0+H_{\rm int}\nonumber\\
&=&\sum_{i=x, y}\epsilon_i {\hat c}_i^\dagger {\hat c}_i
+\frac{1}{2}\sum_{i, j, k, l=x, y}\langle i j|V|k l\rangle
{\hat c}_i^\dagger {\hat c}_j^\dagger {\hat c}_l {\hat c}_k\ ,
\label{ham}
\end{eqnarray}
where $\epsilon_i\equiv \hbar\bar{\omega}+\hbar\omega_i$ ($i=x$, $y$)
and $\bar{\omega}\equiv (\omega_x+\omega_y)/2$,
while ${\hat c}_i^\dagger$ creates and ${\hat c}_i$ destroys a
particle in the state $\psi_i$ \cite{note_xyrep}.  
We shall take the interaction to have
the contact form $\langle {\bf r}, {\bf r}'|V|{\bf r}, {\bf r}'
\rangle =g_{\rm 2D} \delta^2({\bf r}-{\bf r}')$, where $g_{\rm 2D}$ is
an effective two-dimensional interaction strength, which we shall take
to be positive \cite{note_g}.

For orientation we first describe the
GP approach
in which all particles are assumed to be in the
same single-particle state.  Thus the many-body state may be written
as
\begin{align}
|\Delta\phi,\Delta n\rangle \equiv&\frac{1}{\sqrt{N!}}
\left(n_x^{1/2} {\hat c}_x^\dagger
+   n_y^{1/2}  {\hat c}_y^\dagger e^{-i\Delta\phi}\right)^N
|0\rangle.
\label{exphasest}
\end{align}
The quantities $n_x$ and $n_y$ are the occupation probabilities of the
two states, and $n_x+n_y=1$ and  $\Delta n\equiv n_x-n_y$.
The phase difference between the two components is denoted by $\Delta
\phi\equiv \phi_x-\phi_y$,
where $\phi_x$ and $\phi_y$ are the phases of the two states,
and $|0\rangle$ is the
vacuum state.  The GP energy functional is \cite{note_1/N}
\begin{equation}
  \frac{E}{N} = \int d^2r \biggl\{\frac{\hbar^2}{2m}|\nabla\psi|^2
+\frac{m}{2}(\omega_x^2 x^2+\omega_y^2 y^2)|\psi|^2
+\frac{g_{\rm 2D}}{2}|\psi|^4\biggr\} ,
\label{energy}
\end{equation}
where  $\psi$ is the condensate wave function,
$\psi=a_x\psi_x + a_y\psi_y=|a_x|e^{i\phi_x}\psi_x + |a_y|e^{i\phi_y}\psi_y$ with $|a_x|^2+|a_y|^2=N$.
Evaluation of Eq.\ (\ref{energy}) for this wave function leads to
\begin{equation}
{\cal E}\equiv \frac{E}{\hbar\bar{\omega}N}
=\frac{1}{2}\frac{\Delta\omega}{\bar{\omega}} \Delta n
-\frac{\gamma}{4}[1-(\Delta n)^2]\sin{^2\Delta\phi},
\label{enorm}
\end{equation}
where
$\gamma\equiv\gamma_0\sqrt{\omega_x\omega_y/\bar{\omega}^2}$, with
$\gamma_0\equiv Na_{\rm s}/Z$ \cite{note_g}, and
$\Delta\omega \equiv \omega_x-\omega_y$.
In Eq.\ (\ref{enorm}) we have omitted the contribution  independent of
$\Delta n$ and
$\Delta \phi$ \cite{note_coupling}.  One sees that there is one dimensionless
parameter in the problem, $\Gamma\equiv\gamma {\bar\omega}/ {\Delta\omega}$ \cite{distortion}.

\begin{figure}
\rotatebox{0}{
\resizebox{8.2cm}{!}
{\includegraphics{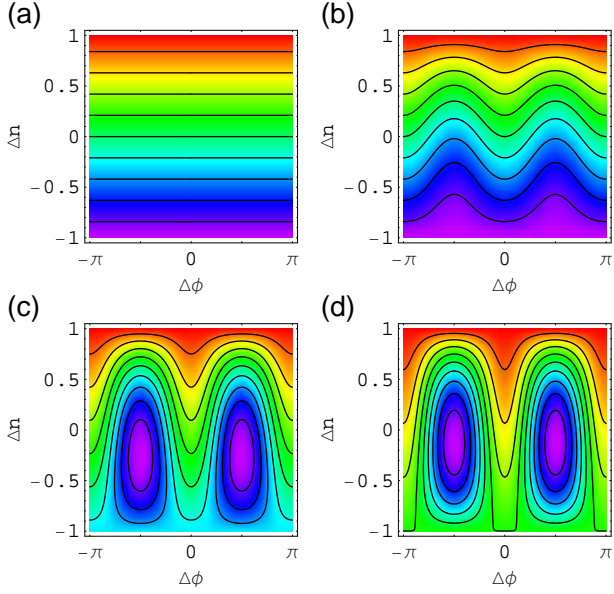}}}
\caption{\label{fig_e}(Color online)
Landscape of the energy per particle ${\cal E}$
for $\Gamma=0$ (a), $0.8$ (b),
$4.0$ (c), and $8.0$ (d).
Purple regions (the darker area for $\Delta n\alt 0$)
correspond to lower ${\cal E}$
and red ones (the darker area for $\Delta n\simeq 1$) to higher ${\cal E}$.
For a given panel, the contour lines are equally spaced in ${\cal E}$,
but the spacing varies from panel to panel.
}
\end{figure}

In Fig.\ \ref{fig_e}, we show contours of ${\cal E}$ as a
function of $\Delta\phi$ and $\Delta n$ for several values of the
interaction strength. States with $\Delta \phi=- \pi/2$ are vortex-like
with positive circulation, while those with   $\Delta \phi=\pi/2 $
correspond to an antivortex, a vortex with negative circulation.  States
with $\Delta \phi=0,\pi$ have nodal lines
which, because of the larger mean square density, give rise to 
a larger repulsive energy.
Allowed motions
correspond to contours of constant energy.  For small $\Gamma$ [Figs.\
\ref{fig_e}(a) and (b)], all
contours are open, and the allowed motions correspond to
oscillations between vortex
states and antivortex ones. For $\Gamma >1$,
global minima of the energy develop
on the lines $\Delta \phi= -\pi/2$ and $\pi/2$, and there are closed
contours surrounding the minima.  These correspond to motions in which a
vortex line oscillates without reversals of the circulation.
With
further increase of $\Gamma$, the closed contours occupy an increasing
fraction of the area [see Figs.\ \ref{fig_e}(c) and (d)] and an
energy barrier with height $\sim \gamma\hbar\bar{\omega}\sim
\hbar\bar{\omega}Na_{\rm s}/Z$ per particle grows between the vortex
and antivortex
 states, i.e., large $\Gamma$
stabilizes the vortex and antivortex states.
Classically, for a system having energy less than the minimum value on
the line $\Delta \phi=0$, the circulation of a vortex cannot change
sign.

\begin{figure}[t]
\rotatebox{270}{
\resizebox{6.5cm}{!}
{\includegraphics{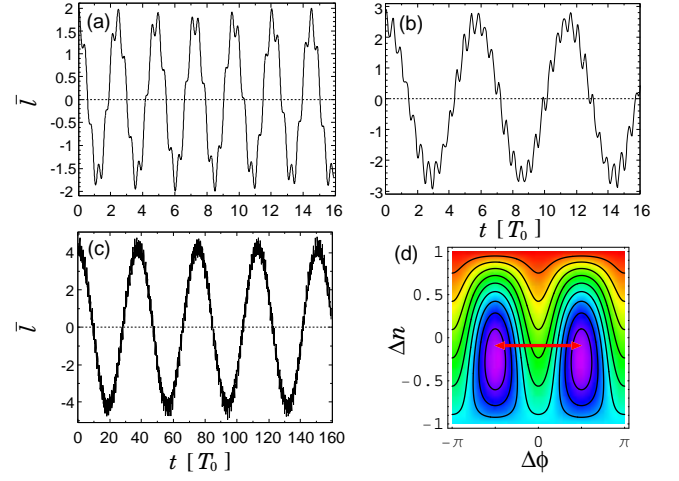}}}
\caption{\label{fig_lt}(Color online)
Time evolution of $\bar{l}$ for $\Gamma=4.0$ and $N=2$ (a), 3 (b), and 5 (c).
Panel (d) is the corresponding energy
landscape showing that the vortex and antivortex states lie on
closed orbits, and the arrow indicates a possible trajectory for tunneling
between these states.}
\end{figure}

However, in quantum mechanics the circulation can reverse by
tunneling.
The interaction Hamiltonian is
given by $H_{\rm int}=(\gamma\hbar{\bar
\omega}/4N)h_{\rm int} $ with
\begin{eqnarray}
h_{\rm int}
&=& {\hat c}_x^\dagger {\hat c}_x^\dagger {\hat c}_y {\hat c}_y
+{\hat c}_y^\dagger {\hat c}_y^\dagger {\hat c}_x {\hat c}_x
+ 4 {\hat c}_x^\dagger {\hat c}_y^\dagger {\hat c}_y {\hat c}_x\nonumber\\
&&+3({\hat c}_x^\dagger {\hat c}_x^\dagger {\hat c}_x
{\hat c}_x
+ {\hat c}_y^\dagger {\hat c}_y^\dagger {\hat c}_y {\hat
c}_y)\nonumber\\
&=&3\hat{N}^2-2\hat{N}-\hat{l\ }^2\ .
\label{hint}
\end{eqnarray}
Here $\hat{N}\equiv {\hat c}_x^\dagger {\hat c}_x + {\hat c}_y^\dagger
{\hat c}_y$
is the total number operator and ${\hat l}\equiv i ({\hat c}_y^\dagger
{\hat c}_x
-
{\hat c}_x^\dagger
{\hat c}_y )  $.  The
expectation value
$L$ of the
$z$-component of the angular momentum operator
in a state
$|\Psi\rangle$ containing only the two single-particle states
$\psi_x$ and
$\psi_y$ is given by
$L= \hbar  \sqrt{\bar{\omega}^2/\omega_x \omega_y}
\langle\Psi|{\hat l}|\Psi\rangle
\equiv \hbar \sqrt{\bar{\omega}^2/\omega_x \omega_y}\ \bar{l}$.
The system has $N+1$ Fock states $|N_x, N_y\rangle=|N_x, N-N_x\rangle$
with $N_x=0,1,\ldots,N$,
where $N_x$ and $N_y$ are the numbers of particles occupying the single
particle states $\psi_x$ and $\psi_y$, respectively.
Using the above many-body Hamiltonian,
we have followed the time evolution of the system, starting from
the vortex state proportional to $({\hat c_x}^\dagger +i{\hat
c_y}^\dagger)^N|0\rangle$,
corresponding to $|-\pi/2, 0\rangle$ for the particular choice
$\Delta\omega/{\bar \omega}=0.01$. Preliminary results were reported in
Ref.\
\cite{lphys}.

Classically, reversals of the circulation with this initial state are
impossible if $\Gamma >2$, which ensures that the state lies on a closed
orbit.
However, we see, e.g., in Fig.\ \ref{fig_lt}
for
$\Gamma=4.0$, that reversals of the angular momentum do occur.
The angular momentum oscillates
with a period
$T\simeq 2.41\, T_0$ for $N=2$,
$T\simeq 5.73\, T_0$ for $N=3$, and
 $T\simeq 37.7\, T_0$ for $N=5$, where $T_0=2\pi/\Delta \omega$ is the
period in the absence of interactions.
The rapid wiggles in Figs.\
\ref{fig_lt}(a)-(c) are due
to oscillations of internal degrees of freedom of the
vortex, which correspond to motion on closed energy contours in the
mean-field approach.

\begin{figure}[htbp]
\rotatebox{270}{
\resizebox{!}{8.5cm}
{\includegraphics{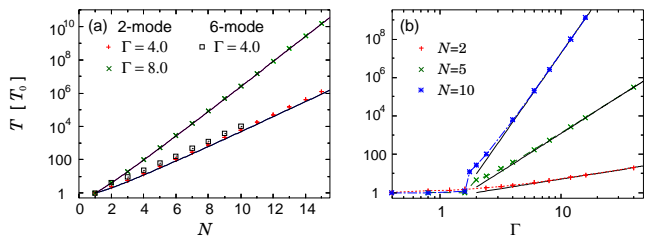}}}
\caption{\label{fig_t}(Color online)
Oscillation period $T$ as a function of $N$
at fixed interaction strength $\Gamma$ (a)
and as a function of $\Gamma$ at fixed $N$ (b).
The thick solid lines for $\Gamma\ge2$ are calculated
from the analytic formula (\ref{deltae}).
The lines connecting the data points in (b) are to guide the eye.
The period $T$ for the data points at $\Gamma=1.6$ in (b)
is hard to define and has an uncertainty of order $0.1\, T_0$.
}
\end{figure}

In Fig.\ \ref{fig_t}(a), we plot the oscillation period $T$ as a
function of $N$ for $\Gamma=4.0$ and $8.0$.
The results do not change
qualitatively for a more general model in which six single-particle
states, the ground state and the three lowest $d$-like states in addition to
the two $p$-like states, are taken into account \cite{note_6st}.  
We see clearly that
$T$ increases almost exponentially with $N$.  This tendency is
consistent with the fact that, for these values of $\Gamma$, the
vortex and antivortex states are stable classically.
We also
see that $T$ increases with $\Gamma$.  This is because for $\Gamma \gg
1$ the Hamiltonian is dominated by $H_{\rm int}$
and therefore level spacings are
proportional to $\gamma$.  Thus mixing of states by the anisotropy
becomes less important for large $\Gamma$.  For $\Gamma<1$, the oscillation
period is approximately $T_0$, as one would expect from the fact that
under these conditions reversals can occur in classical mean-field
theory for all initial conditions.
For $\Gamma\ll 1$, we see
repetitive collapses and revivals of $L$ with a period
$\sim 2\pi N/\gamma\bar{\omega}$ \cite{pitaevskii}.

In Fig.\ \ref{fig_t}(b), we plot $T$ as a function of $\Gamma$ for
fixed $N$ and one sees the different behaviors for $\Gamma
\alt
1$ and
$\Gamma \agt 1$.  The oscillations in the former region are well
described by mean-field theory and $T$ is almost constant with a value
$\sim T_0$.  For large $\Gamma$, vortex-antivortex oscillation occurs for
the given initial conditions only in the quantum-mechanical calculation, and
$T$ shows a power-law dependence on $\Gamma$.

\begin{figure}[htbp]
\rotatebox{270}{
\resizebox{!}{8.5cm}
{\includegraphics{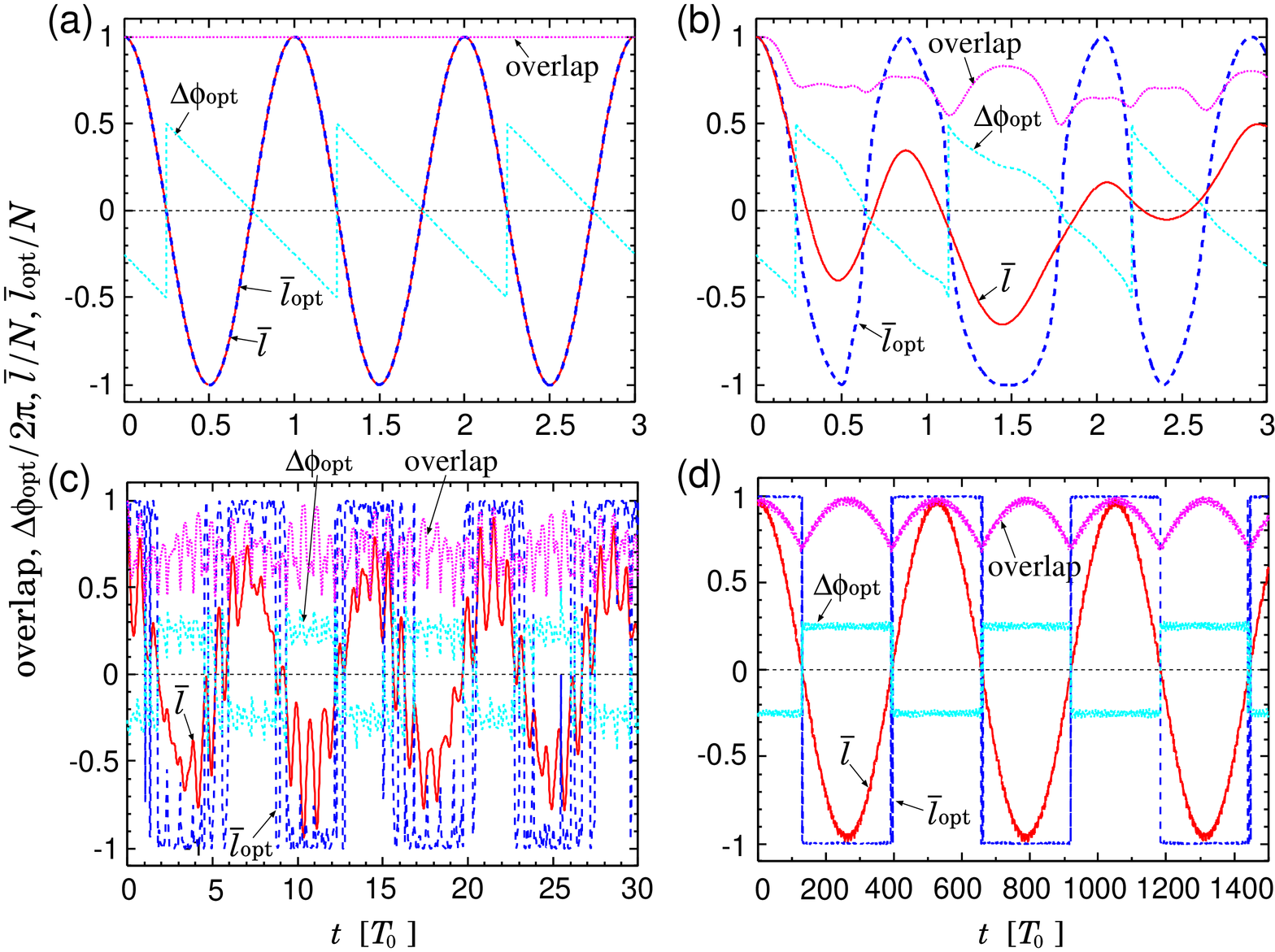}}}
\caption{\label{fig_ovrlp_phase}(Color online)
Overlap $|\langle\Psi|\Delta\phi_{\rm opt},0\rangle|$ of
the wave function with the optimized phase state
for $N=5$ and
$\Gamma=0$ (a), 1.6 (b), 2.4 (c), and 8.0 (d).
}
\vspace{0.5cm}
\rotatebox{0}{
\resizebox{6.cm}{!}
{\includegraphics{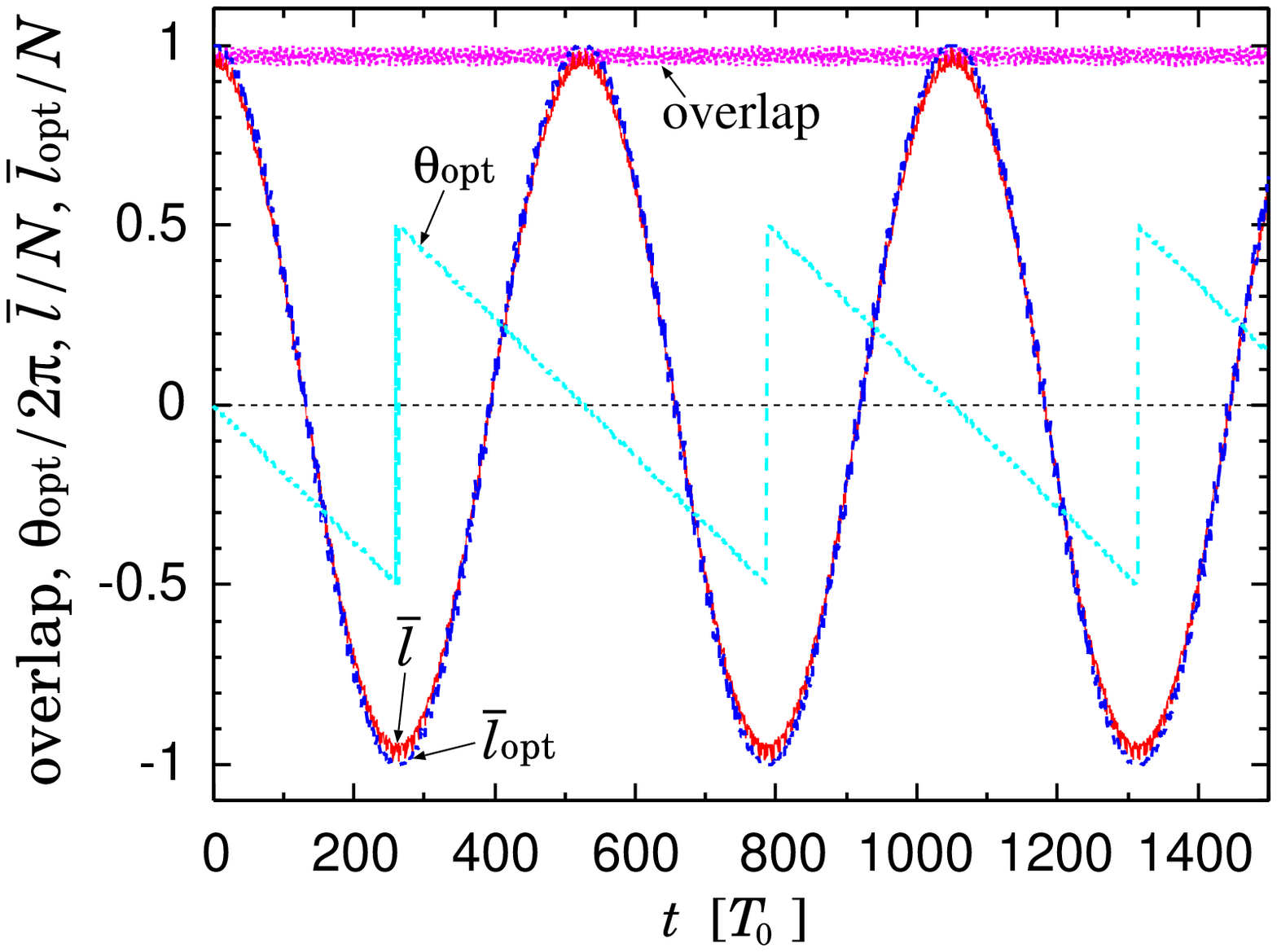}}}
\caption{\label{fig_ovrlp_cat}(Color online)
Overlap $|\langle\Psi|{\rm Cat}; \theta_{\rm opt}\rangle|$ of the numerical
solution of the wave function with the cat state
for $N=5$ and $\Gamma=8.0$.}
\end{figure}

To understand the behavior of the tunneling time,
we have investigated the state $|\Psi(t)\rangle$.
Figure \ref{fig_ovrlp_phase} shows its overlap
$|\langle\Psi|\Delta\phi_{\rm opt},0\rangle|$ with the mean-field state
(\ref{exphasest}),
where $\Delta \phi_{\rm opt}$ is chosen to maximize the overlap.
The expectation value of $\hat{l}$, which is proportional to that of
the angular momentum, is also plotted, and it is denoted by
$\bar{l}$ for $|\Psi(t)\rangle$
and $\bar{l}_{\rm opt}$ for $|\Delta\phi_{\rm opt}, 0\rangle$.
For $\Gamma \alt 1$, $|\Psi(t)\rangle$ is well described by the
mean-field state,  and  $\Delta\phi_{\rm opt}$ changes continuously.
For $\Gamma_0 = 2.4$ and 8.0,  $\Delta\phi_{\rm opt}$ jumps
essentially discontinuously between approximately $-\pi/2$ and $\pi/2$,
corresponding to the vortex and antivortex states. This indicates that
the main components of the state correspond to either all particles
being in the vortex state or all of them in the antivortex state.
A much better approximation for the
wave function for
$\Gamma \gg 1$ is the Schr\"odinger cat state
consisting of a superposition of the states in which all particles are
in the vortex state or all particles are in the antivortex state,
\begin{equation}
  |{\rm Cat}; \theta\rangle
\equiv \cos{\frac{\theta}{2}} \left|-\frac{\pi}{2},0\right\rangle
 + i \sin{\frac{\theta}{2}}
\left|\frac{\pi}{2},0\right\rangle.
\label{cat}
\end{equation}
In Fig.\ \ref{fig_ovrlp_cat}, we plot the overlap
$|\langle\Psi|{\rm Cat}; \theta_{\rm opt}\rangle|$
for the same $|\Psi\rangle$ as in Fig.\ \ref{fig_ovrlp_phase}(d)
\cite{korsbakken}. Similarly, $\theta_{\rm opt}$ is determined by
maximizing
$|\langle\Psi|{\rm Cat}; \theta_{\rm opt}\rangle|$.
Here $L_{\rm opt}$ is the expectation value of the angular momentum in
the state
$|{\rm Cat}; \theta_{\rm opt}\rangle$.
It is remarkable that $\theta_{\rm opt}$ changes almost continuously
and the overlap is close to unity (its mean value is $\simeq 0.97$ for
this case).

The behavior of the wave function for $\Gamma \gg 1$ is most
conveniently analysed in terms of states in which the interaction energy
is diagonal.  These may be written as $|l\rangle=
\left\{ \left[(N+l)/2\right]! \left[(N-l)/2\right]! \right\}^{-1/2}
(\hat{c}_+^\dagger)^{(N+l)/2} (\hat{c}_-^\dagger)^{(N-l)/2}|0\rangle$ where the
operators
${\hat c}_+^\dagger \equiv ({\hat c}_x^\dagger + i{\hat c}_y^\dagger)/\sqrt{2}$
and ${\hat c}_-^\dagger \equiv ({\hat c}_x^\dagger - i{\hat
c}_y^\dagger)/\sqrt{2}$ create particles in a vortex or an antivortex
state, respectively.  The operator $\hat l$ in the subspace of states we
are considering is ${\hat c}_+^\dagger {\hat c}_+ -{\hat c}_-^\dagger
{\hat c}_-$ and the Hamiltonian reduces to
\begin{align}
  H =& \frac{\hbar\Delta\omega}{2} (\hat{c}_+^\dagger \hat{c}_- + \hat{c}_-^\dagger \hat{c}_+)
- \frac{\gamma\hbar\bar{\omega}}{2N} (\hat{c}_+^\dagger \hat{c}_+^\dagger \hat{c}_+ \hat{c}_+ + \hat{c}_-^\dagger \hat{c}_-^\dagger \hat{c}_- \hat{c}_-)\nonumber\\
&-\frac{\gamma\hbar\bar{\omega}}{N} \hat{N}^2.
\label{hvav}
\end{align}
This is equivalent to that of the Bose Hubbard model for two sites if
$+$ and $-$ are regarded as site labels, with hopping matrix element
$-\hbar \Delta \omega/2$ and on-site interaction $-\gamma\hbar {\bar
\omega}/N$ \cite{flach,ho}.
The anisotropy of the trap couples states of different $l$ according
to the Hamiltonian $H'=(\hbar\Delta \omega/2 )({\hat c}_+^\dagger
{\hat c}_- + {\hat c}_-^\dagger {\hat c}_+)$, which we may treat
perturbatively in this regime.  The wave function is dominated by the
vortex and antivortex states, and components from other states are
suppressed by the large interaction energy for these states.
Because of the anisotropy, the vortex and antivortex states
are not energy eigenstates and  the leading contributions to the mixing of
the states for large $\Gamma$ may be calculated by perturbation theory.
Since the anisotropy couples only states in which $l$ differs by 2,
it is of $N$th order in $\Delta \omega$.
The leading contribution to the matrix element mixing the
vortex and antivortex states is
$\Delta E/2 =  H'_{-N,-N+2} (\Delta E_{-N+2})^{-1}\cdots
H'_{N-4,N-2} (\Delta E_{N-2})^{-1} H'_{N-2,N}$,
where $H'_{l,l+2}=\langle l|H'|l+2\rangle$ and
$\Delta E_{l}=E_l-E_{N}$, with
$E_l=-(\gamma\hbar\bar{\omega}/4N) l^2$. 
Here $\Delta E$ is the
splitting of the two lowest states.  For $\Gamma \gg 1$ one finds
\cite{flach}
\begin{equation}
 \Delta E  =\hbar|\Delta\omega| N\left(\frac{\alpha(N)}{\Gamma}\right)^{N-1}\ .
\label{deltae}
\end{equation}
Here $\alpha(N)\equiv N [(N-1)!]^{-1/(N-1)}/2$ is a function  that
depends weakly on $N$: $\alpha(2)=1$ and
$\alpha(\infty)=e/2$. The oscillation period is given by
$T=2\pi\hbar/\Delta E$.
In the strong interaction regime, $\Delta E$ given by Eq.\ (\ref{deltae})
accounts very well for the numerical results, as is shown in Fig.\
\ref{fig_t}.
We remark that rotation of the trap will suppress the vortex-antivortex oscillations
if $2N\hbar\Omega\agt
\Delta E$, where $\Omega$ is the rotational angular velocity.

In conclusion, we have calculated rates for tunneling between vortex and
antivortex states in a simplified model without making
a mean-field approximation and have obtained an analytical result in the
limit of small tunneling.
We have demonstrated that, in this limit, the wave function is much better
described as a Schr\"odinger cat state that is
a superposition of a state in which all particles are in the vortex
state and one in which all particles are in the antivortex state.
We have also shown that the problem is equivalent to a two-site Bose
Hubbard model with attractive interparticle interactions.  On the
experimental side, recent advances in creating few-body
systems trapped on an optical lattice \cite{campbell,foelling} offer the
possibility of observing experimentally the effects we predict.

The authors are grateful to Jason Ho, Jacobus Verbaarschot, and
Augusto Smerzi for valuable discussions and Eugene Zaremba for helpful
comments.  This work was supported by the JSPS 
Postdoctoral Program for Research Abroad.


\begin{thebibliography}{99}
\bibitem{note_anharmonicity} This situation could be realized in practice by,
e.g., making the trap anharmonic, thereby lifting degeneracies present
for a harmonic potential.
\bibitem{note_xyrep} In the representation with $\psi_x$ and $\psi_y$, 
the physical origin of the
flipping phenomenon for weak interactions is clear as we shall
see later.
\bibitem{note_g} For a system uniform in the $z$ direction, and with
the usual Gross-Pitaevskii assumption, $g_{\rm 2D}=g/Z$, where $Z$ is
the extent of the cloud in the $z$ direction, $g\equiv 4\pi\hbar^2
a_{\rm s}/m$ is the two-body interaction strength in three dimensions,
and $a_{\rm s}$ is the $s$-wave scattering length.
\bibitem{note_1/N} We have here made the usual replacement of $1-1/N$
by unity.  This does not affect our conclusions.
\bibitem{note_coupling} Reference \cite{garcia-ripoll} considered the
case $\sin{^2\Delta\phi}=1$, and our Eq.\ (\ref{enorm}) then agrees with
Eq.\ (6) of that paper.
\bibitem{distortion} Distortions of the vortex, which
correspond to the inclusion in the wave function of other
single-particle states, will be unimportant for $\Gamma \alt
|{\bar \omega}/\Delta \omega|$. 
For the numerical calculations we have made, 
this condition is always satisfied.
\bibitem{note_6st} In this case the number of Fock states is 
$(N+1)(N+2)(N+3)(N+4)(N+5) / 120$.
\bibitem{pitaevskii} This can be understood from the fact that
the contribution to the level spacing of the Fock states
from the $\hat{l}^2$ term in $h_{\rm int}$
is $\langle N_x, N_y|\hat{l}^2|N_x, N_y\rangle
=N+\left[N^2-(N_x-N_y)^2\right]/2$,
and thus the situation is essentially the same as that considered
by    A. Imamo\=glu, M. Lewenstein, and L. You,
  Phys.\ Rev.\ Lett.\ {\bf 78}, 2511 (1997) and
  L. P. Pitaevskii,
  Phys.\ Lett.\ A {\bf 229}, 406 (1997).
\end{thebibliography}
\end{document}